\documentclass[runningheads,12pt,a4paper]{llncs}
\usepackage{lineno}
\usepackage{pgf}

\usepackage[centertags]{amsmath}
\usepackage{amsfonts}
\usepackage{ textcomp }
\usepackage{amssymb}
\usepackage{epsfig}
\usepackage{amssymb,latexsym}
\usepackage{indentfirst}

\usepackage[english]{babel}
\usepackage[utf8]{inputenc}
 
\makeatother

\newcommand{\Wa}[1]{\widehat{\chi_{#1}}}

\newcommand{\Tr}{{\rm Tr}}

\newcommand{\boxtensor}{{\Box\kern-9.03pt\raise1.42pt\hbox{$\times$}}}

\newenvironment{pf}{\noindent\textit{Proof.}\quad}{\hfill{$\Box$}}

\newcommand{\sC}{{\mathcal C}}

\newcommand{\C}{{\mathbb C}}

\newcommand{\F}{{\mathbb F}}

\newcommand{\be}{\begin{eqnarray}}
\newcommand{\bd}{\begin{displaymath}}
\newcommand{\ed}{\end{displaymath}}
\newcommand{\ee}{\end{eqnarray}}
\newcommand{\nn}{{\nonumber}}

\usepackage{amssymb}
\usepackage{graphicx}

\usepackage{url}

\urldef{\mailsa}\path|smesnager@univ-paris8.fr, {ozbudak, sahmet}@metu.edu.tr|  
\newcommand{\keywords}[1]{\par\addvspace\baselineskip
\noindent\keywordname\enspace\ignorespaces#1}

\begin{document}

\mainmatter  

\title{A new class of  three-weight linear codes from weakly regular plateaued  functions \footnote {The Extended Abstract of this work was submitted to WCC-2017 (the Tenth International Workshop on Coding and Cryptography).}}

\titlerunning{A new class of  three-weight linear codes  from weakly regular plateaued  functions}

%
%
\author{  Sihem Mesnager\inst{1,2,3} \and Ferruh \"Ozbudak\inst{4,5} \and Ahmet S{\i}nak\inst{2,5,6}}
\authorrunning{Mesnager, \"Ozbudak, S{\i}nak}

\institute{Department of Mathematics, University of Paris VIII, France\\
\and
 LAGA, UMR 7539, CNRS, University of Paris VIII and University of Paris XIII, France\\
\and
Telecom ParisTech, France\\
\and
Department of Mathematics, Middle East Technical University, Turkey\\
\and
Institute of Applied Mathematics, Middle East Technical University, Turkey\\
\and
Department of Mathematics and Computer Sciences, Necmettin Erbakan University, Turkey
\mailsa}

%
%

\toctitle{Lecture Notes in Computer Science}
\tocauthor{Authors' Instructions}
\maketitle

\begin{abstract}
Linear codes with  few weights have many applications in secret sharing schemes, authentication codes, communication and strongly regular graphs. In this paper, we consider linear codes with three weights in arbitrary characteristic. To do this, we generalize the recent contribution of Mesnager given in [Cryptography and Communications 9(1), 71-84,  2017].
 We first  present a new class of binary linear  codes with three weights  from  plateaued  Boolean  functions  and  their weight distributions. 
We next introduce the notion of (weakly) regular plateaued functions in odd characteristic $p$ and give concrete examples of  these functions. Moreover, we construct a new class of three-weight linear $p$-ary codes   from weakly regular plateaued    functions and determine their weight distributions.   We finally analyse the constructed linear codes  for secret sharing schemes.

\end{abstract}

\section{Introduction}
Error correcting codes have many applications in communication systems, data storage devices and consumer electronics.   The construction of linear codes with  few weights has  been widely studied (see, e.g., \cite{DingDingIEEE-Com2014,DingDingIEEE-IT2015,Mesnager_Linear,TangLiQiZhouHelleseth2015,XuCao2015,ZhouLiFanHelleseth2015}) since these codes    have many applications in consumer electronics, secret sharing schemes, authentication codes, communication, data storage system, association schemes, and strongly
regular graphs. 
 Recently,   in \cite{Ding-survey2015}, Ding has published a valuable survey on the construction of binary linear codes from Boolean functions.
 The notion of   plateaued Boolean functions, as an extension of the notion of bent Boolean functions,  has been introduced in \cite{Zeng} by Zheng and Zhang (1999), and then generalized to arbitrary characteristic: the so-called $p$-ary plateaued functions from $\F_{p^n}$ to $\F_p$ (see, e.g., \cite{MesnagerOzbudakSinak-Balkan2015}). Several researchers have  studied plateaued functions since they have  many applications in cryptography, sequence  theory and coding theory. In particular, $p$-ary bent  functions (mostly, quadratic and weakly regular bent functions) have been used in coding theory to construct linear codes with few weights. Very recently,  Mesnager \cite{Mesnager_Linear} has constructed a new family of three-weight linear codes from weakly regular  bent functions in arbitrary characteristic based on a generic construction. Within this framework, the aim of this paper is to construct a class of linear codes with few weights  from weakly regular plateaued functions in arbitrary characteristic and determine their weight distributions.\\
The paper is structured as follows.  Section \ref{preliminaries} sets the main notations and recalls some basic results in coding theory and number theory. In Section \ref{SectionWeakly}, we introduce the notion of (weakly) regular plateaued functions in odd characteristic $p$. We  then give concrete examples to show the existence of   (weakly) regular  plateaued $p$-ary functions. Section \ref{SectionCodes}   constructs a new class of  three-weight linear $p$-ary (resp. binary) codes  from weakly regular $p$-ary plateaued    (resp.  plateaued  Boolean) functions based on a generic construction. We also determine the weight distributions of the constructed linear codes in this paper.  Finally, in Section \ref{SectionSSS},
we   observe that all nonzero codewords of the constructed linear codes are minimal for almost all cases.

\section{Preliminaries}\label{preliminaries}
In this section, we set main notations and give some  basic results on $p$-ary functions,  coding theory  and number theory,  which will be used in the sequel.\\ 
For any set $E$,  $\# E$ denotes the cardinality of $E$ and $E^\star=E\setminus \{0\}$.  
Given a complex number $z\in\mathbb{C}$, $\vert z\vert$ denotes  the absolute value of $z$, where $\mathbb{C}$ is the field of complex numbers.
Let $\F_{p^m}$ be the finite field with $p^m$ elements, where $p$ is a prime and $m\geq 1$ is a positive integer. Then, $\F_{p^m}^{\star}=\langle \zeta \rangle$ is a multiplicative cyclic group of order $p^m-1$ with generator $\zeta$, and $\F_{p}$ is the prime field of $\F_{p^m}$.   The extension field $\mathbb{F}_{p^m}$  can be seen as  an $m$-dimensional vector space over $\F_{p}$, denoted by  $\F_p^m$. 
The absolute trace function $\Tr_{p}^{p^m} : \mathbb {F}_{p^m} \rightarrow \mathbb {F}_{p}$ is defined as
$\Tr_{p}^{p^m}(x):=\sum_{i=0}^{ m-1}x^{p^{i}}.$
Recall that   $\Tr_{p}^{p^m}$ is $\mathbb {F}_{p}$-linear. 
Given a function $f :  \mathbb {F}_{p^m} \longrightarrow  \mathbb {F}_{p}$, the direct and inverse  Walsh transform of $f$ are defined,  respectively, by:
\be\nn
\Wa {f}(b)=\sum_{x\in  \mathbb {F}_{p^m}} {\xi_p}^{{f(x)}-\Tr_{p}^{p^m} (b x)} \mbox{ and }
\ee 
 \begin{equation}\label{inversetransform}
 \xi_p^{f(x)}=\frac{1}{p^m} \sum_{b\in\mathbb {F}_{p^m}} \Wa {f}(b) \xi_p^{\Tr_{p}^{p^m}(bx)},
\end{equation}
where $\xi_p=e^{\frac{2\pi \sqrt {-1}}p}$ is a primitive $p$-th root of unity.
The set $\{b\in \F_{p^m}:  \widehat{\chi_f}(b)\neq 0 \}$ is called the Walsh support of $f$, and is denoted by $Supp\left({\widehat{\chi_f}}\right)$.
\noindent For a nonnegative integer $i$, the moment of Walsh transform of $f$ is  defined by
$
S_i(f)= \sum_{b \in \F_{p^m}} | \widehat{\chi_f}(b)|^{2 i}
$ with the convention  $S_0(f) = p^{m}$, and $S_1(f)=p^{2m}$ is known as  the \textit{Parseval identity}.    Recall that  $f$ is said to be \textit{balanced} over $\F_p$ if $\#\{ x\in \F_p^n : f(x)=k\}=p^{m-1}$  for each $k\in\F_p$, i.e., 
$f$ takes every value of $\F_p$ the same number $p^{m-1}$ times; otherwise, it is called \textit{unbalanced}.\\

\noindent \textbf{Basic background in number theory.} We now recall the basic facts of the Legendre symbol and cyclotomic field.
 Let $a$ be a positive integer and $p$ be an odd prime.  We say that $a$ is a quadratic residue  modulo $p$ if $ \sqrt{a} \in \F_p^{\star}$,  and $a$ is a quadratic non-residue modulo $ p$ if $ \sqrt{a} \notin \F_p^{\star}$. 
The \textit{Legendre symbol} is defined as
\begin{displaymath}
 \left( \frac{a}{p} \right) := \left\{ \begin{array}{ll}
  \,\,\,\, 0 & \textrm{ if }\,  p\mid a,\\
 \,\,\, \, 1 & \textrm{ if $a$ is a quadratic residue  modulo } p,\\
-1 & \textrm { if  $a$ is a quadratic non-residue  modulo } p.
\end{array} \right.
\end{displaymath}
The Legendre symbol satisfies
$
 \left( \frac{a}{p} \right)\equiv a^{\frac{p-1}{2}} \pmod p,
$
and
\be\label{Legendre}   
\left( \frac{-1}{p} \right) \equiv (-1)^{\frac{p-1}{2}}\pmod p= \left\{ \begin{array}{ll}
\,\,\,\, 1 & \textrm{ if}\, \, p \equiv 1 \pmod 4,\\
-1 & \textrm{ if}\, \,  p \equiv 3 \pmod 4.
\end{array} \right.
\ee
Throughout this paper,   $p^*$ denotes $\left(\frac{-1}{p}\right)p$, $\left(\frac{a}{p}\right)$ denotes the  Legendre symbol for $ a \in \F_p^{\star}$,  $\mathbb{Z}$ is the rational integer ring and $\mathbb{Q}$ is the rational field.  
The ring of integers in $\mathbb{Q}(\xi_p)$ is $\mathcal {O}_K:=\mathbb{Z}(\xi_p)$. An integral basis  of $\mathcal {O}_{\mathbb{Q}(\xi_p)}$ is the set  $\{\xi_p^i \mid 1\leq i\leq p-1\}$. The Galois field extension $\mathbb{Q}(\xi_p)/\mathbb{Q}$ of degree $p-1$ is the Galois group $Gal (\mathbb{Q}(\xi_p)/\mathbb{Q})=\{\sigma_a \mid a \in (\mathbb{Z}/p\mathbb{Z})^{\star}\}$, where the automorphism $\sigma_a$ of $\mathbb{Q}(\xi_p)$ is defined by $\sigma_a(\xi_p)=\xi_p^a$. The field $\mathbb{Q}(\xi_p)$ has a unique quadratic subfield $\mathbb{Q}(\sqrt{p^*})$. For $a\in  \F_p^{\star}$, we have $\sigma_a(\sqrt {p^*})=\left(\frac{a}{p}\right)\sqrt{p^*}$. Hence, the Galois group $Gal(\mathbb{Q}(\sqrt{p^*})/\mathbb{Q}=\{1, \sigma_{\gamma}\}$ for any $\gamma \in \F_p$ such that   $\sqrt{\gamma} \notin \F_p^{\star}$.
The reader is referred  to \cite{IrelandRosen1990} for further reading   on cyclotomic fields.\\

\noindent \textbf{Basic background in coding theory.}
\noindent Let $q$ be a prime power and $n$ be a positive integer. The support of a vector $\tilde a=(a_0,\ldots, a_{n-1})\in \mathbb {F}_{q}^n$  is defined as $supp(\tilde a):=\{0 \leq i\leq n-1: a_i\not=0 \}.$ The Hamming weight of  $\tilde a\in \mathbb {F}_{q}^n$,  denoted by $wt(\tilde a)$, is the cardinality of  its support, i.e.,  $wt(\tilde a):=\#supp(\tilde a)$.
A linear $[n,k]_{q}$ code $\mathcal {C}$ over $\mathbb {F}_{q}$ is a $k$-dimensional subspace of $\mathbb {F}_{q}^n$.
A linear $[n,k,d]_{q}$ code $\mathcal {C}$ over $\mathbb {F}_{q}$ is a $k$-dimensional subspace of $\mathbb {F}_{q}^n$ with minimum Hamming distance $d$. 
The dual code of  $\mathcal {C}$ is the linear code with parameters $[n,n-k,d^\perp]_{q}$ defined by 
\be\nn
\mathcal {C}^\perp=\{\tilde b \in\mathbb {F}_{q}^n : \tilde b\cdot \tilde a=\tilde 0 \mbox{ for all } \tilde a\in\sC\},
\ee
where $``\cdot"$ is an inner product in $\mathbb {F}_{q}^n$.
 Let $A_w$ denote the number of codewords with Hamming weight $w$ in $\sC$ of  length $n$. Then, $(1,A_1, \ldots, A_n)$ is the weight distribution of $\sC$ and the polynomial $1+A_1y + \cdots + A_ny^n$ is called the weight enumerator of $\sC$. 
The code  $\sC$ is called a $t$-weight code if the number of nonzero $A_w$ in the weight distribution is  $t$. For further reading on coding theory, we send the reader to \cite{Huffman}.\\

\section{On (weakly) regular plateaued $p$-ary functions}\label{SectionWeakly}
In this section, we introduce the notion of (weakly) regular plateaued functions in odd characteristic $p$ and give some properties of these functions. 
We first recall the notion of plateaued functions. \\
Let $f :\mathbb {F}_{p^m} \longrightarrow  \mathbb {F}_{p}$ be a function.  A $p$-ary function $f$ is called \emph{bent}  if all of its Walsh transform coefficients satisfy  $|\widehat{\chi}_f(b)|^2=p^{m}$, and  $r$-\emph{plateaued} if   all of its Walsh transform coefficients satisfy $ \vert \Wa {f} (b)  \vert^2\in\{0,p^{m+r}\}$,  where $r$  is an integer with $0 \leq r\leq m$. We point out that a $0$-plateaued function  is  bent. In characteristic $2$,  it is safe to say that $f$ is   \emph{  $r$-plateaued Boolean function} if   $\Wa {f}(b) \in\{0,\pm 2^{(m+r)/2}\}$ for all $b\in\F_{2^m}$.
 By the Parseval identity, we  have (see, e.g., \cite{MesnagerOzbudakSinak-Balkan2015}):
\begin{lemma}\label{lemma1} Let $p$ be any prime and $f:\F_{p^m}\to\F_{p}$ be $r$-plateaued. Then for $b \in\F_{p^m}$, $| \widehat{\chi}_f(b)|^2$ takes $p^{m-r}$ times the value $p^{m+r}$ and $p^m-p^{m-r}$ times the value $0$.
\end{lemma}
\begin{lemma}\label{Distribution}
Let $f:\F_{2^m}\to\F_{2}$ be a $r$-plateaued Boolean function. Then for $b \in\F_{2^m}$, 
 the Walsh   distribution  of  $f$ is  given by
\be\nn
\Wa {f}(b)=\left\{\begin{array}{ll}
  2^{\frac{m+r}2}, &   2^{ m-r-1}+ 2^{\frac{m-r-2}2} \mbox{ times, }  \\
0, &   2^{ m }- 2^{m-r} \mbox{ times, }\\
-2^{\frac{m+r}2}, &   2^{ m-r-1}- 2^{\frac{m-r-2}2} \mbox{ times. }\\
     \end{array}\right.
\ee
\end{lemma}
 \noindent   We   recall the notion of (weakly) regular bent functions in odd characteristic $p$ (see, e.g., \cite{HellesethKholosha2006}). For an odd prime $p$, the Walsh transform coefficients of a $p$-ary bent function $f$  
satisfy 
\be\nn
\Wa {f}(b)=\left\{\begin{array}{ll}\pm p^{\frac{m}2} \xi_p^{f^{\star}(b)},  & \mbox{ if } m \mbox{ is even or } m \mbox{ is odd and } p \equiv 1\pmod 4,~ \\
\pm ip^{\frac{m}2} \xi_p^{f^{\star}(b)}, & \mbox{ if } m \mbox{ is odd and } p\equiv 3\pmod 4,\\
     \end{array}\right.
    \ee
where $i$ is a complex primitive $4$-th root of unity and $f^*$ is called the dual of $f$.
 A bent function $f$ is called \emph{regular} if  for  all $b\in \F_{p^m}$, $\Wa {f}(b)=p^{\frac{m}2}\xi_p^{f^*(b)},$ and   \emph{weakly regular} if  there exists a complex number $u$ having unit magnitude (in fact, $|u|=1$ and $u$ does not depend on $b$)
such that 
$
\Wa {f}(b)=up^{\frac{m}2}\xi_p^{f^*(b)}$
 for  all $b\in \F_{p^m}$, where $f^*$ is the dual of $f$; otherwise,  $f$ is called \emph{non-weakly regular}. \\
\noindent Very recently, Hyun et al. \cite{Hyun-Lee-Lee2016} have proved that the Walsh transform  coefficients of a $p$-ary $r$-plateaued function $f$ satisfy
\be\label{PlateauedWalsh}
\Wa {f}(b)=\left\{\begin{array}{ll}
\pm p^{\frac{m+r}2}\xi_p^{g(b)}, 0  & \mbox{ if } m+r \mbox{ is even or } m+r \mbox{ is odd and } p \equiv 1\pmod 4,~ \\
\pm i p^{\frac{m+r}2}\xi_p^{g(b)}, 0 & \mbox{ if } m+r \mbox{ is odd and } p\equiv 3\pmod 4,\\
     \end{array}\right.
      \ee  
where $i$ is a complex primitive $4$-th root of unity and $g$ is  a $p$-ary function over $\mathbb{F}_{p^m}$ with $g(b)=0$ for $b\notin Supp(\Wa {f})$.  
\noindent Notice that by definition of $g:\mathbb{\F}_{p^m}\to \mathbb{\F}_p$, it can be regarded as  a mapping from $Supp(\Wa f)$ to $\mathbb{\F}_p$ since we have $g(b)=0$ for all $b\notin Supp(\Wa f)$.\\
 \noindent The notion of weak regularity is   meaningful for plateaued functions. 
We now introduce the notion of (weakly) regular plateaued functions, which covers a non-trivial subclass of the class of plateaued functions.  
\begin{definition}
Let  $p$ be an odd prime and  $f:\F_{p^m}\to\F_{p}$ be a $p$-ary $r$-plateaued  function, where $r$  is an integer with $0 \leq r\leq m$. Then, $f$ is called \emph{regular $p$-ary $r$-plateaued} if  
$
\Wa {f}(b)\in \{0,p^{\frac{m+r}2}\xi_p^{g(b)} \}
$
for  all $b\in \F_{p^m}$, where $g$ is  a $p$-ary function  over $\mathbb{F}_{p^m}$ with $g(b)=0$ for all $b\notin Supp(\Wa {f})$. Moreover, $f$ is called \emph{weakly regular $p$-ary $r$-plateaued} if  there exists a complex number $u$ having unit magnitude (that is,   $|u|=1$ and $u$ does not depend on $b$) 
such that 
\be\nn
\Wa {f}(b)\in\left\{0, up^{\frac{m+r}2}\xi_p^{g(b)}\right\}
\ee
 for  all $b\in \F_{p^m}$, where  $g$ is  a $p$-ary function  over $\mathbb{F}_{p^m}$ with $g(b)=0$ for all $b\notin Supp(\Wa {f})$; otherwise,  $f$ is called \emph{non-weakly regular $p$-ary $r$-plateaued}.
\end{definition}
\noindent Notice that we have $\Wa {f}(b)=0$ if $b\notin Supp(\Wa {f})$. Then it is safe to  say that    $f$ is  regular  $r$-plateaued if   $\Wa {f}(b)=p^{\frac{m+r}2}\xi_p^{g(b)}$ for  all $b\in Supp(\Wa f)$, and  $f$ is   weakly regular $r$-plateaued if  there exists a complex number $u$ having unit magnitude  
such that 
\be\label{PlateauedWalshh}
\Wa {f}(b)=up^{\frac{m+r}2}\xi_p^{g(b)}
\ee 
 for  all $b\in Supp(\Wa f)$, where  $|u|=1$ (in fact, $u$ can only be equal to $\pm 1$ or $\pm i$ and it does not depend on $b$) and  $g$ is a p-ary function over $Supp(\Wa f)$.
By (\ref{PlateauedWalsh}), regular $r$-plateaued functions can only   exist   for even $m+r$ and for odd $m+r$ with $p\equiv 1\pmod 4$.
 We can derive from  (\ref{PlateauedWalshh}) the following result.
\begin{lemma}\label{WalshFact}
 Let $f$ be a weakly regular  $r$-plateaued   $p$-ary  function. 
For  all $b\in Supp(\Wa f)$, we can say 
$
  \Wa {f}(b)=\epsilon \sqrt{p^*}^{m+r} \xi_p^{g(b)},
$ where $\epsilon=\pm 1$ is    the sign of  $ \Wa f$, $p^*$ denotes $\left(\frac{-1}{p}\right)p$ and  $g$ is a $p$-ary function over $Supp(\Wa f)$. 
\end{lemma}

\begin{pf}
By (\ref{Legendre}) and (\ref{PlateauedWalsh}),   using the fact that $u$ does not depend on $b$ in (\ref{PlateauedWalshh}), we obtain  the following:\\
 If $m+r$ is even or $m+r$ is odd and $p \equiv 1\pmod 4$,  then $\left(\frac{-1}{p}\right)^{m+r}=1$ and $u=\pm 1$ in (\ref{PlateauedWalshh}). Hence, we have $ \epsilon \sqrt{p^*}^{m+r}=\epsilon \sqrt{1}  \sqrt{p}^{m+r}=u \sqrt{p}^{m+r},$ where $ \epsilon=\pm 1$.\\
  If $m+r$ is  odd and $p \equiv 3\pmod 4$, then  $\left(\frac{-1}{p}\right)=-1$ and $u=\epsilon i$ in (\ref{PlateauedWalshh}),  where $ \epsilon=\pm 1$. Hence, $ \epsilon\sqrt{p^*}^{m+r}= \epsilon\sqrt{-1}^{m+r} \sqrt{p}^{m+r}=\epsilon{i}^{m+r} \sqrt{p}^{m+r}= \epsilon{i}\sqrt{p}^{m+r}=u\sqrt{p}^{m+r}.$
The  result now follows.
\end{pf}
  \begin{remark}
 Notice that  the notion of (weakly) regular  $0$-plateaued functions coincides with the one of (weakly) regular  bent functions.  Indeed,  if we have  $|
\Wa {f}(b)|^2\in\{0,p^{m}\}$  for  all $b\in \F_{p^m}$, then by the Parseval identity,  $p^{2m}= p^m \# Supp(\Wa f)$, and so, $ \# Supp(\Wa f)=p^m$. Hence, a (weakly) regular  $0$-plateaued function is the (weakly) regular  bent.
\end{remark}
By MAGMA, we obtain several (weakly) regular $r$-plateaued functions, two of which are given as follows for $p=n=3$.
\begin{example}\label{examplenewodd} 
 A function $f(x)=\Tr_3^{3^3}(\zeta^5x^{11} + \zeta^{20}x^5 + \zeta^{11}x^4 + \zeta^2x^3+\zeta x^2)$ where $\F_{3^3}^{\star}=\langle \zeta \rangle$ with $\zeta^3+2\zeta+1=0$ is regular $3$-ary $1$-plateaued with $\Wa {f}(b)\in\{0,9\xi_3^{g(b)}\}$, where   $g$ is an unbalanced $3$-ary function.
\end{example}
\begin{example}\label{examplenewodd}  
 A function $f(x)=\Tr_3^{3^3}(\zeta x^{13} + \zeta^7x^4 + \zeta^7x^3 +\zeta x^2)$  where $\F_{3^3}^{\star}=\langle \zeta \rangle$ with $\zeta^3+2\zeta+1=0$ is  weakly regular $3$-ary $1$-plateaued with $\Wa {f}(b)\in\{0,-9\xi_3^{g(b)}\}$, where   $g$ is an unbalanced $3$-ary function. On the other hand, a function $\Tr_3^{3^3}(\zeta^{16}x^{13} + \zeta^2x^4 + \zeta^2x^3 + \zeta x^2)$  is  non-weakly regular $3$-ary $2$-plateaued.
\end{example} 

 \noindent The following lemma will be used to determine the weight distributions of the constructed  linear codes. 
\begin{lemma}\label{Walshinverse}
Let  $f$ be a weakly regular  $r$-plateaued $p$-ary   function, that is,  for  all $b\in Supp(\Wa f)$ we have  $\Wa {f}(b)=up^{\frac{m+r}2}\xi_p^{g(b)}$, where  $|u|=1$. Then, we have $$\Wa {g}(x)=u^{-1}p^{\frac{m-r}2}\xi_p^{f(-x)}.$$
\end{lemma}

\begin{pf}
 By the inverse Walsh transform in (\ref{inversetransform}), we have 
\be\nn
\begin{array}{ll}
u^{-1}p^{\frac{m+r}2}\xi_p^{f(x)}&=\displaystyle u^{-1}p^{\frac{m+r}2}\frac{1}{p^m} \sum_{b\in  \F_{p^m}} \Wa {f}(b) \xi_p^{\Tr_{p}^{p^m}(bx)}\\
&=\displaystyle u^{-1}p^{\frac{m+r}2}\frac{1}{p^m} \sum_{b\in Supp(\Wa f)}up^{\frac{m+r}2}\xi_p^{g(b)} \xi_p^{\Tr_{p}^{p^m}(bx)}\\
&=p^r\displaystyle \sum_{b\in Supp(\Wa f)} \xi_p^{g(b)+\Tr_{p}^{p^m}(bx)}=p^r\Wa {g}(-x).
\end{array}
\ee
\end{pf}

\section{A new class of three-weight linear codes from weakly regular plateaued  functions}\label{SectionCodes}
 In this section,  
we construct a new class of linear codes with few weights from plateaued functions in arbitrary characteristic and determine their weight distributions (we shall analyse separately the binary case and the case when $p$ is odd).
 For any $\alpha, \beta\in \mathbb {F}_{p^m}$, one can define a function
\be\nn
\begin{array}{ccccl} f_{\alpha, \beta} &:& \mathbb {F}_{p^m}& \longrightarrow & \mathbb {F}_{p}\\
&&x&\longmapsto& f_{\alpha, \beta}(x):=\Tr_{p}^{p^m}(\alpha \Psi (x)-\beta x),
\end{array}
\ee
where $\Psi$  is a polynomial from $\mathbb {F}_{p^m} $ to  $\mathbb {F}_{p^m}$ such that $\Psi (0)=0$. Then
one can    define a linear code $\mathcal {C}_{\Psi}$ of length $p^m-1$ over $\mathbb {F}_{p}$ as:
\be\nn
 \mathcal {C}_{\Psi}:=\{\tilde {c}_{\alpha, \beta}=(f_{\alpha, \beta}(\zeta_1),f_{\alpha, \beta}(\zeta_2),\ldots, f_{\alpha, \beta}(\zeta_{p^m-1})) \, | \; \alpha,\beta\in\mathbb {F}_{p^m}\},
\ee
where $\zeta_1, \ldots, \zeta_{p^m-1}$ are the  elements of $\mathbb {F}_{p^m}^{\star}$. In this context, the following main results have been obtained in \cite{Mesnager_Linear} by Mesnager.

\begin{proposition} \label{distribution}
 Let $\psi_a$ be a function from $\mathbb {F}_{p^m}$ to $\mathbb {F}_{p}$ defined by
$\psi_a(x)=\Tr_{p}^{p^m}(a\Psi(x))$,
where  $a\in \mathbb {F}_{p^m}$ and $\Psi:\mathbb {F}_{p^m}\to \mathbb {F}_{p^m}$ with $\Psi (0)=0$. For all $ \alpha,\beta\in\mathbb {F}_{p^m}$, we have 
\be\nn
wt (\tilde {c}_{\alpha, \beta})=p^m-\frac {1}p  \sum_{\omega \in\mathbb {F}_{p}} \Wa {\psi_{\omega\alpha}}(\omega \beta).
\ee
\end{proposition}

\noindent We are going to  consider a subclass of the  class of linear codes $ \mathcal {C}_{\Psi}$. We  assume $a=1$ and $\alpha\in \mathbb{F}_p$. Then, we have $f_{\alpha,\beta}(x)=\alpha \psi_1(x)-\Tr_{p}^{p^m}(\beta x)$ and  define a subcode $\mathcal {C}$ of $\mathcal {C}_{\Psi}$ as follows:
\begin{equation}\label{defCode}
\mathcal {C}:=\{\tilde {c}_{\alpha, \beta}=(f_{\alpha, \beta}(\zeta_1), f_{\alpha, \beta}(\zeta_2),\ldots, f_{\alpha, \beta}(\zeta_{p^m-1})) \, | \; \alpha\in\mathbb{F}_{p},\beta\in\mathbb {F}_{p^m}\},
\end{equation}
where $\zeta_1, \ldots, \zeta_{p^m-1}$ are the   elements of $\mathbb {F}_{p^m}^{\star}$.  Then, a linear code $\sC$ over $\F_p$ defined  by $(\ref{defCode})$ is a $k$-dimensional subspace of $\F_p^n$, where $k=m+1$ and $n=p^m-1$, and it is denoted by $[p^m-1,m+1]_p$.
 By  Proposition \ref{distribution}, the Hamming weights of the codewords of $\mathcal {C}$   can be given as follows. 
\begin{proposition} \label{HammingWeight}
We keep  the above arguments.  For $\tilde {c}_{\alpha, \beta}\in \mathcal {C}$,
if $\alpha=0$, we have
$wt (\tilde {c}_{0, 0})=0$ and $wt (\tilde {c}_{0, \beta})=p^m-p^{m-1}$ for $\beta\not=0$, if $\alpha\in\mathbb{F}_p^{\star}$, for all $\beta\in\F_{p^m}$ we have 
\be\nn
wt (\tilde {c}_{\alpha, \beta})=p^m-p^{m-1}-\frac {1}p \sum_{\omega \in\mathbb {F}_{p}^\star} \sigma_\omega\left(\sigma_\alpha(\Wa {\psi_{1}}(\alpha^{-1}\beta))\right),
\ee
where $\alpha^{-1}$ is the multiplicative inverse of $\alpha$ in $\mathbb{F}_p^{\star}$ and     $\sigma_a$  is the automorphism of $\mathbb{Q}(\xi_p)$ for $a \in \F_p^{\star}$.
\end{proposition}

\subsection{A new class of binary three-weight linear codes    from plateaued Boolean  functions}
In this  subsection, we  present  a  new class of binary linear codes with  few weights and their weight distributions using plateaued Boolean   functions.\\
Let  $p=2$ and assume that $\psi_1(x)=\Tr_{2}^{2^m}(\Psi(x))$ is a  $r$-plateaued Boolean  function, where  $m+r$ is even. For  $\alpha\in\F_{2 }$ and $\beta\in\F_{2^m}$, we compute the Hamming weights of the codewords and weight distribution of $\mathcal {C}$ defined  by $(\ref{defCode})$. By Proposition \ref{HammingWeight},   if $\alpha=0$, we have $wt (\tilde {c}_{0, 0})=0$ and $wt (\tilde {c}_{0, \beta})= 2^{m-1}$ for $\beta\not=0$, if  $\alpha=1$ and $\beta\in\F_{2^m}$,  we have
$
wt (\tilde {c}_{1, \beta})= 2^{m-1}-\frac {1}2 \Wa {\psi_{1}}(\beta).
$
By Lemma \ref{Distribution}, 
we have for all $\beta\in\F_{2^m}$,
\be\nn
wt (\tilde {c}_{1, \beta})=\left\{\begin{array}{ll}
 2^{m-1}- 2^{\frac{m+r-2}2}, &   2^{ m-r-1}+ 2^{\frac{m-r-2}2} \mbox{ times, }  \\
2^{m-1}, &   2^{ m }- 2^{m-r} \mbox{ times, }\\
 2^{m-1}+2^{\frac{m+r-2}2}, &   2^{ m-r-1}- 2^{\frac{m-r-2}2} \mbox{ times}.\\
     \end{array}\right.
      \ee

\noindent We   give in the following theorem the Hamming weights of the codewords and the weight distribution of $\sC$. 
\begin{theorem}\label{WTdistributionB}
Let $p=2$ and $\mathcal {C}$ be a binary linear $[2^m-1, m+1]$  code defined  by $(\ref{defCode})$. 
 Assume that  $\psi_1$ is a $r$-plateaued Boolean function, where  $m+r$ is even with $0\leq r\leq m-2$ for $2\leq m$.  Then, the Hamming weight of codewords and  the weight distribution of $\mathcal {C}$ are as in Table \ref{tabloo}.
\begin{table}[!h]
	\begin{center}\label{tabloo}
		\begin{tabular}{|c|c| }
			\hline
			Hamming weight $w$  &  Multiplicity $A_w $  \\ 
			\hline  \hline  
			 0  & 1 \\
			\hline
			$2^{m-1}$  &  $2^{ m+1 }- 2^{m-r}-1$ \\
			\hline
			  $2^{m-1}- 2^{\frac{m+r-2}2}$ &$ 2^{ m-r-1}+ 2^{\frac{m-r-2}2} $\\
			\hline
			  $2^{m-1}+2^{\frac{m+r-2}2}$  &$ 2^{ m-r-1}- 2^{\frac{m-r-2}2} $\\
			\hline
		\end{tabular}
	\end{center}	 \caption{Hamming  weight and multiplicity in $\sC$ when $p=2$ and $m+r$ is even. }	
\end{table}   
\end{theorem}
For $m=5$, a $3$-plateaued Boolean function and the corresponding binary linear code are given.
\begin{example}\label{examplenew}
Let  $\Psi(x)=\zeta^{18}x^5 + \zeta^2x^3$  be a mapping from $\F_{2^5}$ to $\F_{2^5}$, where     $\F_{2^5}^{\star}=\langle \zeta \rangle$ with $\zeta^5+\zeta^2+1=0$. Then,  $\psi_1(x)=\Tr_{2}^{2^5}( \Psi(x))$  is the $3$-plateaued Boolean function, and so  the set $\mathcal {C}$ in $(\ref{defCode})$  is a binary three-wight  linear code with parameters $[31,6]$,  weight enumerator   $1+59y^{16} + 3y^{8} + 1y^{24}$ and  weight distribution $(1,59,3,1)$.
\end{example}

\subsection{A new class of  three-weight linear $p$-ary codes  from  weakly regular plateaued  functions}
  In this subsection,  we  construct a new class of  linear  $p$-ary codes with few weights from weakly regular plateaued  $p$-ary  functions and  determine their weight distributions. \\
\noindent From now on, we assume that $p$ is an odd prime  and  the function $\psi_1(x)=\Tr_{p}^{p^m}(\Psi(x))$ is  weakly regular  $p$-ary $r$-plateaued, where   $r$ is an integer with $0\leq r\leq m$ and  $\Psi:\mathbb {F}_{p^m}\to\mathbb {F}_{p^m}$ with  $\Psi (0)=0$.
 Let $\mathcal {C}$ be a linear $p$-ary code defined by $(\ref{defCode})$ whose codewords are denoted by $\tilde {c}_{\alpha, \beta}$.  
We first compute for all $  \alpha \in\mathbb{F}_p$ and $\beta\in\mathbb{F}_{p^m}$, the Hamming weights of $\tilde {c}_{\alpha, \beta}$   and next determine  the weight distribution of $\sC$.    By Proposition \ref{HammingWeight}, if $\alpha=0$, then we have  $wt (\tilde {c}_{0, 0})=0$ and $wt (\tilde {c}_{0, \beta})=p^m-p^{m-1}$  for $\beta\not=0$.  For   $  \alpha \in\mathbb{F}^{\star}_p$,  to compute $wt(\tilde {c}_{\alpha, \beta})$, we need the following.

\begin{lemma}\label{NewLemma}
Let $f:\F_{p^m}\to\F_{p}$ be $r$-plateaued, where $r$  is an integer with $0 \leq r \leq m$.  Define the sets 
$W:=\{(\alpha,\beta)\in\mathbb{F}^{\star}_p\times\mathbb{F}_{p^m}\mid  \widehat{\chi}_f({\alpha^{-1}}\beta)=0\}$
and 
\be\nn\label{Set}
WS:=\{(\alpha,\beta)\in\mathbb{F}^{\star}_p\times\mathbb{F}_{p^m}\mid  \widehat{\chi}_f({\alpha^{-1}}\beta)\neq 0\}.
\ee
Then, the cardinalities of $W$ and $WS$ are equal respectively to $(p-1)(p^m-p^{m-r})$ and   $(p-1)p^{m-r}$.
\end{lemma}

\begin{pf}
By Lemma \ref{lemma1}, we have $\#\{\beta\in\mathbb{F}_{p^m}\mid  \Wa {f}(\beta)=0\}=p^m-p^{m-r}$ and $\#Supp( \Wa {f})=p^{m-r}$, where $Supp( \Wa {f})=\{\beta\in\mathbb{F}_{p^m}\mid  \Wa {f}(\beta)\neq0\}$. Hence, the   result follows.
\end{pf}\\

\noindent For all $  \alpha \in\mathbb{F}^{\star}_p$ and $\beta\in\F_{p^m}$,  by Proposition \ref{HammingWeight},  we have  
\be\label{weightdistt}
wt (\tilde {c}_{\alpha, \beta})=p^m-p^{m-1}-\frac {1}p \sum_{\omega \in\mathbb {F}_{p}^\star} \sigma_\omega\left(\sigma_\alpha(\Wa {\psi_{1}}(\alpha^{-1}\beta))\right).
\ee
 Then there are two cases: $\Wa {\psi_1}( \alpha^{-1}\beta)=0$ or $\ne 0$. 
If  $ (\alpha, \beta)\in W$, i.e., $\Wa {\psi_1}( \alpha^{-1}\beta)=0$, then  we have
$
wt (\tilde {c}_{\alpha, \beta})=p^m-p^{m-1},
$
 that is,  the number of codewords of Hamming weight $p^m-p^{m-1}$ is equal to the cardinality of  $W$  by Lemma \ref{NewLemma}. If $ (\alpha,\beta)\in WS$, i.e., $\Wa {\psi_1}( \alpha^{-1}\beta)\neq 0$, to compute $wt (\tilde {c}_{\alpha, \beta})$ in (\ref{weightdistt}),   we  use the following    (see Lemma \ref{WalshFact})
\be\nn
\Wa {\psi_1}( \alpha^{-1}\beta)=\epsilon \sqrt{p^*}^{m+r}\xi_p^{g( \alpha^{-1}\beta)},
\ee
 where $\epsilon=\pm 1$, $p^*$ denotes $\left(\frac{-1}{p}\right)p$ and  $g$ is a p-ary function over $Supp(\Wa {\psi_1})$.    Notice that 
we have $\sigma_\alpha(\sqrt {p^*}^{m+r})=\sigma_\alpha( \sqrt p^*)^{m+r}=\left(\frac{\alpha}{p}\right)^{m+r}\sqrt {p^*}^{m+r}$, where $\sigma_\alpha$  is the automorphism of $\mathbb{Q}(\xi_p)$ for $\alpha \in \F_p^{\star}$.  
Then we get
\be\nn
\begin{array}{ll}
\vspace{.3 cm}
 & \sigma_\omega\left(\sigma_\alpha(\Wa {\psi_1}( \alpha^{-1} \beta))\right)=\sigma_\omega\Big(\epsilon \left({\frac{\alpha}{p}}\right)^{m+r}\sqrt {p^*}^{m+r} \xi_p^{\alpha g( \alpha^{-1}\beta)}\Big)=\\

&\epsilon\left(\frac{\alpha}{p}\right)^{m+r}\sigma_\omega(\sqrt {p^*}^{m+r}) \xi_p^{\omega\alpha g( \alpha^{-1}\beta)}=\epsilon\left({\frac{\alpha}{p}}\right)^{m+r}\left(\frac{\omega}{p}\right)^{m+r}\sqrt {p^*}^{m+r} \xi_p^{\omega\alpha g( \alpha^{-1}\beta)},
\end{array}
\ee
where  $\sigma_\omega$ is  the automorphism of $\mathbb{Q}(\xi_p)$ for $\omega \in \F_p^{\star}$.
Notice that $\left(\frac{a}p\right)^{m+r}=1$ and $\sqrt {p^*}^{m+r}=\sqrt {p}^{m+r}$  if $m+r$ is even; otherwise, $\left(\frac{a}p\right)^{m+r}=\left(\frac{a}p\right)$ for $a\in\F_p^{\star}$.
Hence, by (\ref{weightdistt}) we have
\be\nn
wt (\tilde {c}_{\alpha, \beta})=\left\{
\begin{array}{ll}\label{Distributionofm}
p^m-p^{m-1}-\epsilon\frac{1}p\left(\frac{\alpha}{p}\right)\sqrt {p^*}^{m+r}\sum_{\omega\in\mathbb{F}^{\star}_p}\left(\frac{\omega}{p}\right) \xi_p^{\omega\alpha g( \alpha^{-1}\beta)},
&\mbox { if } m+r \mbox { odd, } \\
p^m-p^{m-1}-\epsilon p^{\frac{m+r}2-1}\sum_{\omega\in\mathbb{F}^{\star}_p}
 \xi_p^{\omega\alpha g( \alpha^{-1}\beta)}, &\mbox { if } m+r \mbox { even.}\nn
\end{array}\nn
\right.
\ee
We now investigate two   cases.  First, assume $m+r$ odd.
  If $g( \alpha^{-1}\beta)=0$, then 
\be\nn
wt (\tilde {c}_{\alpha, \beta})&=p^m-p^{m-1}-\epsilon\frac{1}p\left(\frac{\alpha}{p}\right)\sqrt {p^*}^{m+r} \displaystyle\sum_{\omega\in\mathbb{F}^{\star}_p}\left(\frac{\omega}{p}\right)=p^m-p^{m-1},
\ee
where  we used $ \sum_{\omega \in\mathbb{F}^{\star}_p} \left(\frac{\omega}{p}\right)=0$. If $g( \alpha^{-1}\beta)\not=0$, then we have
\be\nn
&\displaystyle \sum_{\omega\in\mathbb{F}^{\star}_p}\left(\frac{\omega}{p}\right) (\xi_p^{\omega})^{\alpha g( \alpha^{-1}\beta)}=  \sigma_{\alpha g( \alpha^{-1}\beta)}\left(\sum_{\omega\in\mathbb{F}^{\star}_p}\left(\frac{\omega}{p}\right) \xi_p^{\omega}\right)=  \sigma_{\alpha g( \alpha^{-1}\beta)}(\sqrt {p^*}) =  \left(\frac{\alpha g( \alpha^{-1}\beta)}{p}\right)\sqrt {p^*},
\ee
where we used    $\sum_{\omega\in\mathbb{F}^{\star}_p}(\frac{\omega}{p}) \xi_p^{\omega}=\sqrt {p^*}$. 
Hence, 
\be\nn
\begin{array}{ll}
wt (\tilde {c}_{\alpha, \beta})&=p^m-p^{m-1}-\epsilon\frac{1}{p} \sqrt{p^*}^{m+r+1} \left(\frac{\alpha^2}{p}\right) \left(\frac{ g( \alpha^{-1}\beta)}{p}\right)\\
&=p^m-p^{m-1}-\epsilon\frac{1}{p} \left(\frac{-1}{p}\right)^{\frac{m+r+1}2}p^{\frac{m+r+1}2} \left(\frac{ g( \alpha^{-1}\beta)}{p}\right)\\
&=p^m-p^{m-1}-\epsilon  \left(-1\right)^{\frac{(p-1)(m+r+1)}{4}}p^{\frac{m+r-1}2} \left(\frac{ g( \alpha^{-1}\beta)}{p}\right),
\end{array}
\ee
where we used $\left(\frac{\alpha}{p}\right)\left(\frac{\alpha}{p}\right)=\left(\frac{\alpha^2}{p}\right)=1$, $p^*=\left(\frac{-1}{p}\right)p$ and  $\left(\frac{-1}{p}\right)=(-1)^{\frac{p-1}{2}}$.\\  
Now,  assume $m+r$  even.  If $g( \alpha^{-1}\beta)=0$, then we have
\begin{displaymath}
 wt (\tilde {c}_{\alpha, \beta})=p^m-p^{m-1}-\epsilon p^{\frac{m+r-2}2}(p-1);
\end{displaymath}
otherwise, we have
$
wt (\tilde {c}_{\alpha, \beta})=p^m-p^{m-1}+\epsilon p^{\frac{m+r-2}2}
$
because  if $g( \alpha^{-1}\beta)\not=0$, then
$\sum_{\omega\in \mathbb{F}^{\star}_p}\xi_p^{\alpha\omega g(\alpha^{-1}\beta)}=-1$ since   $\sum_{j=0}^{p-1}x^j$  is the minimal polynomial of $\xi_p$ over $\mathbb{Q}$.  \\

\noindent We now collect in the following theorem the Hamming weights of the codewords of $\mathcal {C}$  defined by $(\ref{defCode})$.
\begin{theorem}\label{WTdistribution}
Let $\mathcal {C}$ be a linear $p$-ary code defined by $(\ref{defCode})$.  Assume that   $\psi_1$ is   weakly regular $p$-ary $r$-plateaued.  Then,  for all $  \alpha \in\mathbb{F}_p$ and $\beta\in\mathbb{F}_{p^m}$, the Hamming weights of $\tilde {c}_{\alpha, \beta}$ are given as follows.
For $\alpha=0$, we have $wt (\tilde {c}_{0, 0})=0$ and $wt (\tilde {c}_{0, \beta})=p^m-p^{m-1}$ for $\beta\not=0$.  For $\alpha\in\F_p^{\star}$ and $\beta\in\F_{p^m}$,  if $ (\alpha, \beta)\in W$, i.e., $\Wa {\psi_1}( \alpha^{-1}\beta)=0$, then
 we get $wt (\tilde {c}_{\alpha, \beta})=p^m-p^{m-1}$, and if $ (\alpha,\beta)\in WS$ , i.e., $\Wa {\psi_1}( \alpha^{-1}\beta)\neq 0$,  then
 
\begin{itemize}
\item when $m+r$ is odd,  
\begin{displaymath}
wt (\tilde {c}_{\alpha, \beta})=\left\{
\begin{array}{ll} 
p^m-p^{m-1},
\mbox { if } \alpha\in\mathbb{F}_{p}^\star$ \mbox { and } $g( \alpha^{-1}\beta)=0, \\
p^m-p^{m-1}-\epsilon \left(-1\right)^{\frac{(p-1)(m+r+1)}{4}}p^{\frac{m+r-1}2} \left(\frac{ g( \alpha^{-1}\beta)}{p}\right),
 \mbox { if } \alpha, g( \alpha^{-1}\beta)\in\mathbb{F}_{p}^\star,\\
 \end{array}
\right.\\
\end{displaymath}

\item when $m+r$ is even, 
\begin{displaymath}
wt (\tilde {c}_{\alpha, \beta})=\left\{
\begin{array}{ll} 
p^m-p^{m-1}-\epsilon(p-1)p^{\frac{m+r-2}2},
\mbox { if } \alpha\in\mathbb{F}_{p}^\star$ \mbox { and } $g( \alpha^{-1}\beta)=0, \\
p^m-p^{m-1}+\epsilon p^{\frac{m+r-2}2},
 \mbox { if } \alpha, g(\alpha^{-1}\beta)\in\mathbb{F}_{p}^\star,\\
 \end{array}
\right.\\
\end{displaymath}
 where $\epsilon=\pm 1$ is     the sign of  $ \Wa {\psi_1}$.
\end{itemize}
\end{theorem}

\noindent Now we are going to determine the weight distributions of  the constructed codes given in Theorem \ref{WTdistribution}.  To do this, we first give the following result. 
  By Lemma \ref{Walshinverse}, the Walsh transform of $g$ is written as
 \be\nn
\Wa {g}(x)=\epsilon vp^{\frac{m-r}2}\xi_p^{ \psi_1(-x)},
\ee
where $\epsilon=\pm 1$ denotes the sign of $\Wa{g}$ and $v\in\{1,i\}$ in $\C$. 
By using this    for $x=0$, we can compute the number of $b\in Supp(\Wa { \psi_1})$ such that  $g(b)=j$ for all $j\in\F_{p}$. 
Set 
$$N_{g}(j):=\#\{b\in Supp(\Wa { \psi_1}) \mid g(b)=j\}.$$
Notice that  $g(b)=0$ for all $b\notin Supp(\Wa { \psi_1})$, and $\#Supp(\Wa { \psi_1})=p^{m-r}.$ 
Hence, we have 
\be\label{Numberg}
\sum_{j=0}^{p-1} N_{g}(j)=p^{m-r}.
\ee 
 
\begin{remark} \label{Balanced}
If $g$ is balanced over $Supp(\Wa { \psi_1})$,   we have
$N_{g}(j)=p^{m-r-1}$ for all $j\in\F_{p}$.
\end{remark}
We include the proof of the following proposition for making the paper self-contained (see, e.g.,  \cite{HellesethKholosha2013,Mesnager_Linear}).
\begin{proposition}\label{Prop:annexe}
We keep  the above notations and assume that $g$ is unbalanced over $Supp(\Wa { \psi_1})$. Then we have the following.  If $m-r$ is even, then
\be\nn
N_{g}(j)=
\left\{\begin{array}{ll}
p^{m-r-1}+\epsilon p^{\frac{m-r-2}2}(p-1),&  j=0,\\
p^{m-r-1}-\epsilon p^{\frac{m-r-2}2},&    j\in\F_p^{\star}.
\end{array}\right.
\ee
 If $m-r$ is odd, then
\be\nn
N_{g}(j)=
\left\{\begin{array}{ll}
p^{m-r-1}, & j=0,\\
p^{m-r-1}+\epsilon p^{\frac{m-r-1}2}\left(\frac{j}{p}\right), & j\in\F_p^{\star},
\end{array}\right.
\ee
where $\epsilon=\pm 1$ is the sign of $\Wa{g}$.
\end{proposition}

\begin{pf}
 Using the Walsh value of unbalanced  $g$ at point zero, then we have
 $$\Wa g(0)=\sum_{b\in Supp(\Wa { \psi_1}) }\xi^{g(b)}_p=\sum_{j=0}^{p-1}N_{g}(j)\xi^j_p=\epsilon vp^{\frac{m-r}2}\xi^{\psi_1(0)}_p$$
equivalently,
\begin{equation}\label{equation1}
\sum_{j=0}^{p-1}N_{g}(j)\xi^j_p-\epsilon vp^{\frac{m-r}2}=0.
\end{equation}
  If $m-r$ is even, then $v=1$. Because $\sum_{j=0}^{p-1}x^j$  is the minimal polynomial of $\xi_p$ over the rational number field, then for all $ j\in\F_p^{\star}$ we have 
\be\nn
N_{g}(j)=a, \mbox{ and } N_{g}(0)=a+\epsilon p^{\frac{m-r}2}
\ee
for some constant $a$. 
 By (\ref{Numberg}), $a+\epsilon p^{\frac{m-r}2}+(p-1)a=p^{m-r}$ from which one deduces $a=p^{m-r-1}-\epsilon p^{\frac{m-r}2-1}$.\\
  If $m-r$ is odd, then
$
v=\left\{\begin{array}{ll}1,  & \mbox{ if }  p \equiv 1\pmod 4, \\
i ,& \mbox{ if } p\equiv 3\pmod 4.\\
     \end{array}\right.$\\
 Recall the well-known identity (see, e.g.,\cite{LidlNiederreter1997})
\be\nn
\sum_{j=0}^{p-1}\left(\frac{j}p\right)\xi^j_p=\left\{\begin{array}{ll}p^{\frac{1}2};  & \mbox{ if }  p \equiv 1\pmod 4,~ \\
ip^{\frac{1}2},& \mbox{ if } p\equiv 3\pmod 4,\\
     \end{array}\right.
    \ee
    that is, $\sum_{j=0}^{p-1}\left(\frac{j}p\right)\xi^j_p=vp^{\frac {1}2}$.	Thus, (\ref {equation1}) can  be rewritten as
   
\be\nn
\sum_{j=0}^{p-1}N_{g}(j)\xi^j_p-\epsilon p^{\frac{m-r-1}2}\sum_{j=0}^{p-1} \left(\frac {j}p\right)\xi^j_p=0;
\ee
equivalently,
\be\nn
\sum_{j=0}^{p-1}\xi^j_p\left(N_{g}(j)-\epsilon p^{\frac{m-r-1}2}  \left(\frac {j}p\right)\right)=0.
\ee
 Then for all $ j\in\F_p^{\star}$, we have
$
N_{g}(j)=N_{g}(0)+\epsilon p^{\frac{m-r-1}2}\left(\frac{j}p\right).
$ By (\ref{Numberg}),  we obtain $\sum_{j=0}^{p-1} N_{g}(j)=pN_{g}(0)+\epsilon p^{\frac{m-r-1}2}\sum_{j=0}^{p-1} \left(\frac{j}p\right)=p^{m-r}$. Thus, since $\sum_{j=0}^{p-1} \left(\frac{j}p\right)=0$, the proof is complete.
\end{pf}\\

\noindent We  can derive from Remark \ref{Balanced} and  Proposition \ref{Prop:annexe} the weight distributions of  the constructed codes.
\begin{theorem}\label{weight-even}
Let $\mathcal {C}$ be a linear $p$-ary code defined by $(\ref{defCode})$.
Assume that   $\psi_1$ is   weakly regular $p$-ary $r$-plateaued and $m+r$ is even with $0\leq r\leq m-2$ for $2\leq m$.  Then, the Hamming weights of codewords and  the weight distributions of $[p^m-1,m+1]$ code $\cal{C}$  are as in Tables \ref{table11} and  \ref{table1} if  $g$ is unbalanced and  balanced over $Supp(\Wa { \psi_1})$, respectively, where  $\epsilon=\pm 1$ is the sign of $\Wa{\psi_1}$.

\begin{table}[!h]
\begin{center}	
\begin{tabular}{|c|c|c|}
\hline
Hamming weight  $w$  & Multiplicity $A_w$  \\

\hline
\hline
$0$ & $1$\\
\hline
 {$p^m-p^{m-1}$} &  {$p^{m+1}-p^{m-r}(p-1)-1$}\\
\hline
 {$p^m-p^{m-1}-\epsilon(p-1)p^{\frac{m+r-2}2}$} &  $p^{m-r-1}(p-1)+\epsilon p^{\frac{m-r-2}2}(p-1)^2$\\
\hline
 {$p^m-p^{m-1}+\epsilon p^{\frac{m+r-2}2}$} &  $(p^{m-r}-p^{m-r-1})(p-1)-\epsilon p^{\frac{m-r-2}2}(p-1)^2$\\
\hline
\end{tabular}
\end{center}	
\caption{
\label{table11} Hamming  weight and multiplicity in $\mathcal{C}$ when $m+r$ is even and $p$ is odd for unbalanced $g$}
 
\begin{center}	
\begin{tabular}{|c|c|c|}
\hline
Hamming weight $w$   & Multiplicity $A_w$ \\
\hline
\hline
$0$ & $1$\\
\hline
 {$p^m-p^{m-1}$} &  {$p^{m+1}-p^{m-r}(p-1)-1$}\\
\hline
 {$p^m-p^{m-1}-\epsilon(p-1)p^{\frac{m+r-2}2}$} &  $p^{m-r-1}(p-1)$\\
\hline
 {$p^m-p^{m-1}+\epsilon p^{\frac{m+r-2}2}$} &  $\left(p^{m-r}-p^{m-r-1}\right)(p-1)$\\
\hline
\end{tabular} \end{center}	
\caption{
\label{table1} Hamming  weight and multiplicity in $\mathcal{C}$ when $m+r$ is even and $p$ is odd for balanced $g$}
\end{table}
\end{theorem}
\begin{pf}
By Theorem \ref{WTdistribution}, the numbers of codewords of Hamming weight $0$ and of Hamming weight  $p^m-p^{m-1}$ are equal respectively to $1$ and  $p^m-1+\# W=p^{m+1}+p^{m-r}-p^{m-r+1}-1$. Now  we are going to determine the weight distribution of  $\mathcal{C}$ for $(\alpha,\beta)\in WS$, i.e., $\Wa{ \psi_1}( \alpha^{-1}\beta)\neq 0$.
Set 
\be\nn
\begin{array}{ll}
N_g(0)&:=\#\{\gamma\in Supp(\Wa { \psi_1})\mid g(\gamma)=0\},\\ K_g(0)&:=\#\{(\alpha,\beta)\in\mathbb{F}^{\star}_p\times\mathbb{F}_{p^m}\mid g({\alpha^{-1}}\beta)=0\},\\
KS_g&:=\#\{(\alpha,\beta)\in\mathbb{F}^{\star}_p\times\mathbb{F}_{p^m}\mid g({\alpha^{-1}}\beta)\neq 0\}.
\end{array}
\ee 
Notice that  for all $b\notin Supp(\Wa { \psi_1})$,   $g(b)=0$ by definition of $g$ and so,  $g( \alpha^{-1}\beta)=0$   for all $ (\alpha, \beta)\in W$. Hence,   by Lemma \ref{NewLemma}, $K_g(0)=\# W+(p-1) N_g(0)$ and
 $KS_g=(p-1)p^m-K_g(0)$.  Assume that $g$ is unbalanced over $Supp(\Wa { \psi_1})$.
 Then, since $N_g(0)=p^{m-r-1}+\epsilon p^{(m-r-2)/2}(p-1)$ by Proposition \ref{Prop:annexe}, we have
\begin{displaymath}
K_g(0)=\# W+p^{m-r-1}(p-1)+\epsilon p^{\frac{m-r-2}2}(p-1)^2,\\
\end{displaymath}
and
$KS_g=(p^{m-r}-p^{m-r-1})(p-1)-\epsilon p^{(m-r-2)/2}(p-1)^2.$ 
Hence, by Theorem \ref{WTdistribution},  the numbers of codewords of Hamming weight $p^{m}-p^{m-1}-\epsilon (p-1)p^{(m+r-2)/2}$  and of Hamming weight $p^m-p^{m-1}+\epsilon  p^{(m+r-2)/2}$ are equal  to  $K_g(0)-\# W$ and  $KS_g$, respectively.\\ 
 Assume that $g$ is balanced over $Supp(\Wa { \psi_1})$.
By Remark \ref{Balanced}, $N_g(0)=p^{m-r-1}$, and so we have $K_g(0)=\# W+p^{m-r-1}(p-1)$
and $KS_g=\left(p^{m-r}-p^{m-r-1}\right)(p-1)$.  As in the first case, the assertion holds.
\end{pf}\\

\noindent For  $p=3$ and $m=3$,  a weakly regular $3$-ary $1$-plateaued function and the corresponding   linear $3$-ary code are given as follows.
 \begin{example}\label{ }
Let  $\Psi:\F_{3^3}\to \F_{3^3}$ be a map defined by $\Psi(x)=\zeta^{22}x^{13} + \zeta^7x^4 + \zeta x^2$  where $\F_{3^3}^{\star}=\langle \zeta \rangle$ with $\zeta^3+2\zeta+1=0$. A function $\psi_1(x)=\Tr_{3}^{3^3}( \Psi(x))$  is weakly regular $3$-ary $1$-plateaued with $\Wa { \psi_1}(b)\in\{0,-9\xi_3^{g(b)}\}$, where   $g$ is an unbalanced $3$-ary function.  Then, the set $\mathcal {C}$ in $(\ref{defCode})$  is a three-wight   linear $3$-ary code with parameters $[26,4]_3$,  weight enumerator   $1+62y^{18} +2y^{24} + 16y^{15}$ and  weight distribution $(1,62,2,16)$.
\end{example}

\begin{theorem}\label{weight-odd} Let $\mathcal {C}$ be a linear $p$-ary code defined by $(\ref{defCode})$. Assume that   $\psi_1$ is   weakly regular $p$-ary $r$-plateaued and  $m+r$ is odd with $0\leq r\leq m-1$.  Then, the Hamming weights of codewords and the weight distributions of  $[p^m-1,m+1]$    code $\cal{C}$  are as  in Tables \ref{table22} and  \ref{table2} if  $g$ is unbalanced  and balanced over $Supp(\Wa { \psi_1})$, respectively, where  $\epsilon=\pm 1$ is the sign of $\Wa{\psi_1}$. 
\begin{table}[!]
\begin{center}	
\begin{tabular}{|c|c|c|}
\hline
Hamming weight $w$   & Multiplicity $A_w$ \\
\hline
\hline
\footnotesize{$0$} & \footnotesize{$1$}\\
\hline
\footnotesize{$p^m-p^{m-1}$} & \footnotesize{$p^{m+1}-p^{m-r-1} (p-1)^2-1$}\\
\hline
\footnotesize{$p^m-p^{m-1}-\epsilon \left(-1\right)^{\frac{(p-1)(m+r+1)}{4}} p^{\frac{m+r-1}{2}}$} & \footnotesize{$\frac{1}{2}(p^{m-r-1}+\epsilon p^{\frac{m-r-1}2})(p-1)^2$}\\
\hline
\footnotesize{$p^m-p^{m-1}+\epsilon\left(-1\right)^{\frac{(p-1)(m+r+1)}{4}} p^{\frac{m+r-1}{2}}$} & \footnotesize{$\frac{1}{2}(p^{m-r-1}-\epsilon p^{\frac{m-r-1}2})(p-1)^2$}\\
\hline
\end{tabular}\end{center}	
\caption{
\label{table22} Hamming  weight and multiplicity in $\mathcal{C}$ when $m+r$ and $p$ are odd  for unbalanced $g$}
\begin{center}	
\begin{tabular}{|c|c|c|}
\hline
Hamming weight $w$   & Multiplicity $A_w$ \\
\hline
\hline

\footnotesize{$0$} & \footnotesize{$1$}\\
\hline
\footnotesize{$p^m-p^{m-1}$} & \footnotesize{$p^{m+1}-p^{m-r-1} (p-1)^2-1$}\\

\hline
\footnotesize{$p^m-p^{m-1}-\epsilon \left(-1\right)^{\frac{(p-1)(m+r+1)}{4}} p^{\frac{m+r-1}{2}}$} & \footnotesize{$\frac{1}{2}p^{m-r-1} (p-1)^2 $}\\

\hline

\footnotesize{$p^m-p^{m-1}+\epsilon \left(-1\right)^{\frac{(p-1)(m+r+1)}{4}} p^{\frac{m+r-1}{2}}$} & \footnotesize{$\frac{1}{2}p^{m-r-1}(p-1)^2$}\\
\hline
\end{tabular}\end{center}	
\caption{
 \label{table2} Hamming  weight and multiplicity in  $\mathcal{C}$ when $m+r$ and $p$ are odd for balanced $g$}
\end{table}

\end{theorem}
\begin{pf} 
Set $N_g(j):=\#\{\gamma\in Supp(\Wa { \psi_1})\mid g(\gamma)=j\}$ and $K_g(j):=\#\{(\alpha,\beta)\in\mathbb{F}^{\star}_p\times\mathbb{F}_{p^m}\mid g({\alpha^{-1}}\beta)=j\}$ for all $j\in\F_p$. Notice that  for all $b\notin Supp(\Wa { \psi_1})$,   $g(b)=0$ by definition of $g$ and so,    $g( \alpha^{-1}\beta)=0$ for all $ (\alpha, \beta)\in W$. Then,   by Lemma \ref{NewLemma}, $K_g(0)=\# W+(p-1) N_g(0)$ where $N_g(0)=p^{m-r-1}$  (see Remark \ref{Balanced} and Proposition \ref{Prop:annexe}). Hence, by Theorem \ref{WTdistribution}, the number of codewords of Hamming weight $p^m-p^{m-1}$ is equal to $p^m-1+K_g(0)=p^{m+1}+2p^{m-r}-p^{m-r+1}-p^{m-r-1}-1.$
 Moreover,
the number of codewords of Hamming weight $p^m-p^{m-1}-\epsilon \left(-1\right)^{(p-1)(m+r+1)/4}p^{(m+r-1)/2}$ and of Hamming weight $p^m-p^{m-1}+\epsilon \left(-1\right)^{(p-1)(m+r+1)/4}p^{(m+r-1)/2}$ is equal respectively to $\sum_{j\in\{1,\ldots, p-1\}, \left(\frac{j}p\right)=1}(p-1) N_g(j)$ and $\sum_{j\in\{1,\ldots, p-1\}, \left(\frac{j}p\right)=-1} (p-1) N_g(j)$. 
If $g$ is unbalanced, then  by Proposition \ref{Prop:annexe},  
\be\nn
\begin{array}{ll}
\displaystyle \sum_{j\in\{1,\ldots, p-1\}, \left(\frac{j}p\right)=1}(p-1)N_g(j)&=\displaystyle \sum_{j\in\{1,\ldots, p-1\}, \left(\frac{j}p\right)=1}(p-1)(p^{m-r-1}+\epsilon p^{\frac{m-r-1}2})\\
&=\frac {(p-1)^2}2(p^{m-r-1}+\epsilon p^{\frac{m-r-1}2})\\
\end{array}
\ee
and
\be\nn
\begin{array}{ll}
\displaystyle \sum_{j\in\{1,\ldots, p-1\}, \left(\frac{j}p\right)=-1}(p-1)N_g(j)&=\displaystyle\sum_{j\in\{1,\ldots, p-1\}, \left(\frac{j}p\right)=-1}(p-1) (p^{m-r-1}-\epsilon p^{\frac{m-r-1}2})\\
&=\frac {(p-1)^2}2(p^{m-r-1}-\epsilon p^{\frac{m-r-1}2}).\\
\end{array}
\ee
If $g$ is balanced then by Remark \ref{Balanced},  
$\displaystyle \sum_{j\in\{1,\ldots, p-1\}, \left(\frac{j}p\right)=1}(p-1)N_g(j)=\frac {(p-1)^2}2p^{m-r-1} $ and \\
$\displaystyle \sum_{j\in\{1,\ldots, p-1\}, \left(\frac{j}p\right)=-1}(p-1)N_g(j)=\frac {(p-1)^2}2p^{m-r-1}.$ The proof is complete.
\end{pf}

%
%
  \begin{remark}We finally should remark that if we assume only the weakly regular bentness in this paper, then we can recover the results given  in   \cite{Mesnager_Linear}  by Mesnager. Therefore, this paper can be viewed as an extension of  \cite{Mesnager_Linear} to the notion of weakly regular  $r$-plateaued functions for any positive integer $r$.
\end{remark}

\section{The constructed three-weight  linear codes for secret sharing schemes}\label{SectionSSS}

In this section, we consider  our linear codes presented in Section \ref{SectionCodes} for secret sharing schemes. A linear code provides a pair of secret sharing schemes, based on a linear code $\sC$ and its dual code $\sC^\perp$. For the secret sharing scheme based on the dual code  $\sC^\perp$, we need to find all minimal codewords of $\sC$. 
We say that a vector $\tilde a$   covers a vector $\tilde b$ if  $supp(\tilde b)\subset supp(\tilde a) $. Then, if a nonzero codeword $\tilde a$  of $\sC$ does not cover any other nonzero codeword of $\sC$, then $\tilde a$ is called  \textit{minimal codeword} of $\sC$. The \textit{covering problem} is to find all the minimal codewords of $\sC$. In general, this problem is very hard, but it can be easy for some  linear codes. Then the main question is how to find a linear code whose  all  nonzero codewords  are minimal. For more details, we send the reader to \cite{DingDingIEEE-IT2015}.
\begin{lemma}\label{Minimality}\cite{Ashikhmin}
Let $\sC$ be a linear code over $\F_p$. Every nonzero codeword of $\sC$ is minimal if 
$\frac{p-1}{p}<\frac{w_{min}}{w_{max}},
$
where $w_{min}$ and $w_{max}$ denote the minimum and maximum nonzero weights in $\sC$, respectively.

\end{lemma}

\noindent We now consider the constructed linear codes   in Theorems \ref{WTdistributionB}, \ref{weight-even}  and  \ref{weight-odd}.
Let $\mathcal {C}$ be the binary linear  code  of Theorem \ref{WTdistributionB} and  $m+r$ be even. Then we readily see that 
$  \frac{1}{2} <\frac{w_{min}}{w_{max}},$
where $w_{min}=2^{m-1}- 2^{(m+r-2)/2}$ and $w_{max}=2^{m-1}+2^{(m+r-2)/2}$
since we have $3\cdot 2^{(m+r)/2}< 2^{m}$ for $m\geq 4$ and  $0\leq r\leq m-4$.
Hence, by Lemma \ref{Minimality},
all  nonzero codewords of  $\mathcal {C}$ given in Theorem \ref{WTdistributionB}   are minimal  if $m\geq 4$ and  $0\leq r\leq m-4$.\\

\noindent Let $p$ be any odd prime, $m+r$ be even and $\mathcal {C}$ be the linear $p$-ary     code of Theorem \ref{weight-even}.  If $\epsilon=1$, 
we have $w_{min}=p^m-p^{m-1}- (p-1)p^{(m+r-2)/2}$ and $w_{max}=p^m-p^{m-1}+p^{(m+r-2)/2}$. If  $\epsilon=-1$,
then $w_{min}=p^m-p^{m-1}- p^{(m+r-2)/2}$ and $w_{max}=p^m-p^{m-1}+(p-1)p^{(m+r-2)/2}$. 
For both cases, we see  that
$ \frac{p-1}{p} <\frac{w_{min}}{w_{max}}$ for $m\geq 4$ and  $0\leq r\leq m-4$ 
 since we have $(p+1) p^{(m+r)/2}<p^{m}$ if $\epsilon=1$ and  $(p^2-p+1) p^{(m+r)/2}<p^{m}(p-1)$ if $\epsilon=-1$. 
 Hence,  by Lemma \ref{Minimality}, all  nonzero codewords of   $\mathcal {C}$ given in  Theorem \ref{weight-even}  are minimal if $m\geq 4$ and  $0\leq r\leq m-4$. \\
 
\noindent Let $p$ be any odd prime, $m+r$ be odd and $\mathcal {C}$ be the linear   $p$-ary   code  of Theorem \ref{weight-odd}.  Then we    see that 
$ \frac{p-1}{p} <\frac{w_{min}}{w_{max}},$
where $w_{min}=p^m-p^{m-1}- p^{(m+r-1)/2}$ and $w_{max}=p^m-p^{m-1}+p^{(m+r-1)/2}$
since we have $(2p-1)p^{(m+r+1)/2}< p^{m}(p-1)$ for $m\geq 3$ and  $0\leq r\leq m-3$.
 Hence,  by Lemma \ref{Minimality},
all  nonzero codewords of     $\mathcal {C}$ given in Theorem \ref{weight-odd}  are minimal if $m\geq 3$ and  $0\leq r\leq m-3$. 

 \section{Conclusion}
The paper studies for the first time constructions of linear codes with few weights from   weakly regular plateaued functions. We   first present a new family of binary three-weight linear codes    from plateaued Boolean   functions and   their weight distributions.  In odd characteristic $p$,
we  introduce the notion of (weakly) regular plateaued functions and give concrete examples of these functions. We next present a new family of three-weight linear $p$-ary codes  from weakly regular plateaued    functions, and   their weight distributions.  We finally analyse the constructed linear codes in this paper   for secret sharing schemes. The constructed linear codes are inequivalent to the known ones (since there is no code with the obtained parameters) in literature as far as we know.
\section*{Acknowledgment} 
The third author is  supported by TÜBİTAK (the Scientific and Technological Research Council of Turkey), program no: BİDEB 2214/A.

\end{document}